\documentclass[twocolumn]{autart}  
\usepackage{graphicx} 

\usepackage[caption=false,font=normalsize,labelfont=sf,textfont=sf]{subfig}
\usepackage{cite}

\usepackage{amsmath,amsfonts,amssymb}
\usepackage{xcolor}
\usepackage{tikz}
\usetikzlibrary{arrows.meta,    
                calc, chains,   
                positioning,    
                quotes          
                }

\usepackage{algpseudocode}
\usepackage{algorithm}
\newtheorem{theorem}{Theorem}
\newtheorem{remark}{Remark}
\newtheorem{assumption}{Assumption}
\newtheorem{definition}{Definition}
\newtheorem{proposition}{Proposition}

\newcommand{\voutput}{\widetilde{y}}
\newcommand{\voutputclean}{y}

\newcommand{\vinput}{u}
\newcommand{\vpinput}{\widetilde{u}}

\newcommand{\K}{k}

\newcommand{\ts}{t}
\newcommand{\vd}{d}
\newcommand{\vr}{r}
\newcommand{\ve}{e}
\newcommand{\para}{\theta}
\newcommand{\paraest}{\widehat{\theta}}
\newcommand{\regressor}{\varphi}

\newcommand{\grad}{\psi}
\newcommand{\gradest}{\widehat{\psi}}
\newcommand{\yd}{\delta}

\newcommand{\constb}{h}
\newcommand{\constA}{\widetilde{g}_{n_d+1}}

\newcommand{\ydmax}{\yd_{\max}}
\newcommand{\ydmin}{\yd_{\min}}
\newcommand{\dmax}{d_{\max}}
\newcommand{\dmin}{d_{\min}}

\newcommand{\noisev}{\lambda}
\newcommand{\noisevest}{\widehat{\noisev}}

\newcommand{\armaxtext}{\textsc{armax}}
\newcommand{\prbstext}{\textsc{prbs}}
\newcommand{\rpemtext}{\textsc{rpem}}
\newcommand{\E}{\mathbb{E}}

\newcommand{\paraexp}{\vartheta}
\newcommand{\paraexpest}{\widehat{\vartheta}}
\newcommand{\orderexcit}{n_e}
\newcommand{\orderdelay}{n_d}

\begin{document}
\begin{frontmatter}
\title{
Recursive Experiment Design for Closed-Loop Identification of ARMAX Systems with Output Perturbation Limits 
}

\author[uu-it]{Jingwei Hu,}
\author[uu-it]{Dave Zachariah,} 
\author[uu-it]{Torbj\"orn Wigren,} 
\author[uu-it]{Petre Stoica}
\address[uu-it]{Division of Systems and Control, Department of Information Technology,
Uppsala University, Sweden.}  


\begin{abstract}
In many applications, system identification experiments must be performed in closed loop to ensure safety or to maintain system operation. In this paper, we consider the recursive design of informative experiments for ARMAX models by adding a bounded probing signal to the input  generated by a fixed output feedback controller. The resulting output perturbations should be kept within user-specified limits. We analyze the identifiability and feasibility conditions of this setting and then proceed to derive a probing signal that can be efficiently computed in closed form. We demonstrate the effectiveness and properties of the design in numerical experiments.
\end{abstract}
\end{frontmatter}

\section{Introduction}

%
%

The design of informative experiments is of central importance in system identification \cite{goodwin1977dynamic,soderstrom1989system,ljung1999system}. To maintain the system in safe operation it is often necessary to perform experiments in \emph{closed-loop} \cite{forssell1999closed}. In this setting, informativeness is determined by the closed-loop dynamics, as well as by the excitation available in the measured signals. In some closed-loop configurations, structural factors such as controller dynamics, plant/controller delays, and specific degree relations in the plant-controller interconnection may render the input-output data informative for the identification of certain model classes, even when the controller reference is constant. When such structural richness is insufficient, additional exogenous excitation is required
\cite{gustavsson1977identification,van1992delay,gevers2008informative,shardt2011closed}.

One approach to obtain informative closed-loop data is to shape the closed-loop
structure itself, by designing a controller that excites the system while
steering it within operational constraints \cite{astrom2008adaptive}. Several receding-horizon control methods have been developed to incorporate a variety of performance objectives and constraints \cite{rathousky2013mpc,marafioti2014persistently,heirung2015mpc,larsson2015experimental,hildebrand2015closed,hu2025adaptive}. Since the system model parameters are unknown and continuously estimated, a major challenge of these adaptive control methods is to ensure good closed-loop performance during the experiment.

In many practical cases, however, there is already a known output feedback controller in place. Then the purpose of the estimated system model is often to improve the existing controller and the design of experiments is achieved by perturbing either the set-point or control signal. This has been tackled using frequency-based methods \cite{bombois2004least,bazanella2010closed,boukhebouz2021shaping}, where the designs are to be synthesized in the time-domain. However, these fixed designs require a good initial estimate of the system model and do not adapt to the information obtained sequentially during the experiment.


In this paper, we propose an alternative time-based experiment design method for closed-loop systems, operating with a known output feedback controller by adding a probing signal to the input signal. Our contributions include:
\begin{itemize}
  \item An adaptive and \emph{recursive} design of a probing signal that uses sequentially estimated model parameters from a recursive prediction error method  \cite{ljung1999system}.
  \item A computationally efficient \emph{closed-form} solution for minimization of the estimation error covariance, while adaptively steering the system within specified limits on the perturbed output.
  \item An analysis of the conditions ensuring that the experimental data are informative enough for closed-loop identification and that the specified perturbation limits are recursively feasible. 
\end{itemize}
We demonstrate that the method yields accurate identification of the system parameters, while effectively steering the system output within specified perturbation limits.

\begin{figure}
    \centering
   \resizebox{0.9\linewidth}{!}{%
        \tikzstyle{block} = [draw, rectangle, 
    minimum height=3em, minimum width=4em]

\tikzstyle{sum} = [draw, circle, inner sep=0pt, minimum size=1em, node distance=1cm]
\tikzstyle{input} = [coordinate]
\tikzstyle{output} = [coordinate]
\tikzstyle{pinstyle} = [pin edge={to-,thin,black}]

\begin{tikzpicture}[auto, node distance=1.5cm,>=latex]
    \node [input, name=input] {};
    \node [sum, right of=input] (sum) {$+$};
    \node [block, right of=sum] (controller) {$K$};
    \node [input, right of=controller, above of=controller, node distance = 1.5cm] (dist_input) {};
    \node [sum, right of=controller, node distance = 1.5cm] (dist_sum) {$+$};
    \node [block, right of=dist_sum,
            node distance=1.5cm] (system) {$G$};
 
    \draw [->] (controller) -- node[name=u] {$u$} (dist_sum);
    \draw [->] (dist_input) -- node[name=d] {$\textcolor{red}{\vd}$}  (dist_sum);
    \draw [->] (dist_sum) -- (system);
    \node [sum, right of=system, node distance=1.5cm] (noise_sum) {$+$};
    \node [block, above of=noise_sum] (noise) {$H$};
    \node [input, left of=noise] (noise_input){$e$};
    \node [output, right of=noise_sum] (output) {};

    \draw [draw,->] (input) -- node {$r$} (sum);
    \draw [->] (sum) -- node {$ $} (controller);
    \draw [->] (system) -- (noise_sum);
    \draw [->] (noise_sum) -- node [name=y,pos=.7] { $\voutput = \voutputclean + \textcolor{red}{\yd}$}(output);
    \draw [->] (noise) -- (noise_sum);
    \draw [draw, ->] (noise_input) -- node {$e$} (noise);
    \node[input, below of=dist_sum, node distance = 1.5cm](feedback){};
    
    \draw[-] (y) |- (feedback)
    node[midway,above]{};

    \draw[-latex] (feedback)-|  (sum)
    node[right,pos=0.9]{$-$};
\end{tikzpicture}
    }
    \caption{Closed-loop system where $G$ and $H$ are unknown and $K$ is a known output feedback controller with a set-point $r$.  A \emph{probing} signal $\vd$ is added and the resulting output is $\voutput = \voutputclean + \yd$, where $\voutputclean$ is the nominal (unperturbed) output and $\yd$ is the \emph{perturbation}. We seek a recursive design of  $\vd$ that yields information about $G$ and $H$ while limiting $\yd$.}
    \label{fig:closedloop}
    \vspace{-1em}
\end{figure}

The rest of the paper is outlined as follows. First, we formulate the problem of recursive experiment design for \armaxtext{} systems in closed-loop operation. Then we provide an analysis of identifiability and recursive feasibility conditions. Following this analysis, we propose a design method that is closed-form and satisfies the derived identifiability conditions. Finally, the recursive designs are evaluated in numerical experiments, illustrating their ability to probe the system of interest while steering the resulting perturbations within operational limits.

\emph{Notation:} $q^{-1}$ is the backward shift operator, such that $q^{-1}x_t = x_{t-1}$. We define $( x )_+ = \max\{0, x\}$. The variance of a random variable is denoted $\mathbb{V}$. $P \mid Q$ signifies that
the polynomial $P$ divides the polynomial $Q$.

\section{Problem formulation}

We want to identify a model of a linear dynamical system
\begin{equation}
y_t = G(q;\para) u_t + H(q; \para) e_t,
\label{eq:linearsystem}
\end{equation}
while in operation where $y_\ts$ is the output, $\vinput_\ts$ is a control input, and $\ve_\ts$ is the
innovation process. The following assumption specifies the
distributional properties of the innovation.
\begin{assumption}[Innovation distribution]\label{ass:innovation-distribution}
The sequence $\ve_\ts$  is a zero-mean white-noise signal with variance
$\noisev>0$ and a non-degenerate continuous distribution.
\end{assumption}

We consider the widely used \armaxtext{} structure \cite{ljung1999system}, so that
\begin{equation}
    G(q;\para) = \frac{q^{-\orderdelay}B(q;\para)}{A(q;\para)} \quad \text{and} \quad  H(q;\para) = \frac{C(q;\para)}{A(q;\para)},
\label{eq:armax}
\end{equation}
where 
\begin{equation*}
\begin{split}
A(q;\para) &= 1 + \sum_{i=1}^{n_a}a_iq^{-i},
\\ B(q;\para) &= \sum_{i=1}^{n_b}b_iq^{-i}, \text{ and} \\
C(q;\para) &= 1 + \sum_{i=1}^{n_c}c_iq^{-i}.
\end{split}
\end{equation*}
The model is therefore parameterized by:
\begin{equation}
    \theta = [b_1 \: \cdots \: b_{n_b} \: a_1 \: \cdots \: a_{n_a} \: c_1 \: \cdots \: c_{n_c}]^\top \label{armax},
\end{equation}
where
$n_a$, $n_b$ and $n_c$ are the model orders, and $\orderdelay$ the input delay (i.e. the integer delay beyond the inherent one-step delay induced by the strict causality of $B(q;\theta)$). The  \emph{aim} is to recursively estimate $\para$ with a prediction error method using sequentially collected input-output data  \cite{ljung1999system,soderstrom1989system}.

We restrict our attention to regular \armaxtext{} parametrizations for which
the following condition holds.
\begin{assumption}[ARMAX regularity]\label{ass:armax-regularity}
The polynomials $A(q;\theta)$, $B(q;\theta)$, and $C(q;\theta)$ have no
common factor. Moreover, $C(q;\theta)$ is invertible.
\end{assumption}

The system is operated in closed loop with a known linear feedback
controller,\begin{equation}
\vinput_\ts = K(q)(\vr_\ts - \voutputclean_\ts),
\label{eq:linearcontroller}
\end{equation}
where $r_t$ is a known external reference signal. This configuration is
illustrated in Fig.~\ref{fig:closedloop}. 
We assume that the feedback  controller  is well behaved in
the following sense.
\begin{assumption}[Closed-loop stability]\label{ass:closed-loop-stability}
The closed-loop system induced by the known controller $K(q)$ is
exponentially stable \cite{forssell1999closed}.
\end{assumption}
A central aim closed-loop identification is then to estimate an accurate system model that can be used to improve a controller.


The input-output data in closed-loop operation may not be sufficiently informative to identify the system. We therefore consider adding a \emph{probing signal} \(\vd_{\ts} \in[\dmin, \dmax]\) to the input, i.e.,
\begin{equation}
\vpinput_\ts = \vinput_\ts + \vd_\ts.
\label{eq:perturbedinput}
\end{equation}
with the aim of obtaining informative data. However, injecting a non-zero $\vd_t$ into the linear closed-loop system with delay results in a perturbed output:
\begin{equation*}
\voutput_{\ts+\orderdelay+1} = \voutputclean_{\ts+\orderdelay+1} + \yd_{\ts+\orderdelay+1},
\end{equation*}
where $\voutputclean_{\ts+\orderdelay+1}$ represents the nominal (unperturbed) output and  $\yd_{\ts+\orderdelay+1}$ is a \emph{perturbation} that we consider to be experimental cost. The goal is to \emph{recursively design} the perturbation $\vd_{\ts}$ so that the resulting process $\{ (\vpinput_\ts,\voutput_{\ts} ) \}$ is informative, while steering the perturbation within user-specified limits
\begin{equation}
\ydmin \leq \yd_{\ts+\orderdelay+1} \leq \ydmax,
\label{eq:performanceconst}
\end{equation}
which are application dependent. (For the zero perturbation to be admissible, we naturally require $\ydmin\le0\le\ydmax$.)

Given the information available at each $\ts$, we pose the design problem for $\vd_\ts$ as minimizing some measure of the resulting parameter errors, denoted $J_{\ts+1}$, under the specified perturbation limits:
\begin{equation}
    \begin{split}
         \min_{\vd_\ts} & \: J_{\ts+1} \\
         \text{s.t. } & \: \dmin \leq \vd_\ts \leq \dmax\\
         & \: \ydmin \leq \yd_{\ts+i} \leq \ydmax,\;i=1,2,\dots.\\
         \end{split}
         \label{eq:optim_ideal}
\end{equation}
We will consider two design criteria $J_{\ts+1}$ in Section~\ref{sec:method}. This is a challenging problem since it involves multi-step constraints.


\section{Analysis}
In this section, we analyze the probing signal $\vd_\ts$ from two perspectives: informativeness and feasibility. We first study conditions that ensure the experimental input-output data is sufficiently informative for identifying the \armaxtext{} system in closed-loop. Then we turn to examining conditions under which the multi-step perturbation limits in \eqref{eq:optim_ideal} are  recursively feasible.

\subsection{Informativeness of input-output data}

To prepare for the analysis of $\{ (\vpinput_\ts,\voutput_{\ts} ) \}$,  we represent the known controller as 
\begin{equation}\label{eq:controller_polynomial}
K(q) = \frac{L(q)}{M(q)},    
\end{equation}
where 
\begin{equation*}
L(q) = \ell_0 + \sum_{i=1}^{n_\ell}\ell_i q^{-i} \text{ and } M(q) = 1 + \sum_{i=1}^{n_m}m_i q^{-i}.
\end{equation*}
We impose the following standard coprimeness condition on this
factorization
\begin{assumption}[Controller coprimeness]\label{ass:controller-coprime}
The polynomials $L(q)$ and $M(q)$ are coprime.
\end{assumption}

The first result re-state conditions that ensure $\{ (\vpinput_\ts,\voutput_{\ts} ) \}$ is informative \emph{without} any probing signal, see also \cite[ch. C10.1]{soderstrom1989system}\cite{gevers2008informative}. We subsequently consider the use of a probing signal $\vd_{\ts}$.

\begin{theorem} \label{thm:ealone}
Consider a closed-loop experiment with constant reference ($r_t\equiv 0$) and no added probing signal $(d_t\equiv 0$). Define the integer
\begin{equation}
n_p\triangleq\min\!\bigl(n_c,\max(n_a+n_m,\;n_b+n_\ell+\orderdelay)\bigr).
\label{eq:np_upper}
\end{equation}
If 
\begin{equation}
\gamma \triangleq n_p+\min\!\bigl(n_a-n_\ell-\orderdelay,\;n_b-n_m\bigr)< 0 ,
\label{eq:lambdacondition}
\end{equation}
then $\{ (\vpinput_\ts,\voutput_{\ts} ) \}$ is informative enough to identify the \armaxtext{} model. Thus \eqref{eq:lambdacondition} is a sufficient condition.
\end{theorem}
\begin{pf}
Following the approach in \cite{ljung1999system}, we consider the linear one-step predictor
\[
\hat{y}_{t}
=
\underbrace{H(q;\theta)^{-1}G(q;\theta)}_{W_u(q;\theta)}(u_t+d_t)
+
\underbrace{\bigl(1-H(q;\theta)^{-1}\bigr)}_{W_y(q;\theta)} y_t .
\] 
which forms an equivalent parameterization of the system model. Now consider two parameters $\para_1$ and $\para_2$, which yield $(A_1, B_1, C_1)$ and $(A_2, B_2, C_2)$, respectively. The difference between the two corresponding predictors is then
\begin{equation}
\Delta \hat{y}_t
\triangleq
\Delta W_u(q;\theta)(u_t+d_t)
+
\Delta W_y(q;\theta)y_t,\label{eq:prediction_diff}
\end{equation}
where
\begin{equation}
\label{eq:deltaWs}
\begin{split}
\Delta W_u \triangleq \frac{q^{-\orderdelay}B_1}{C_1} - \frac{q^{-\orderdelay}B_2}{C_2} \quad \text{and} \quad
\Delta W_y \triangleq \frac{A_2}{C_2}-\frac{A_1}{C_1} 
\end{split}    
\end{equation}
using \eqref{eq:armax} and omitting $q$ for notational simplicity. We assume $d_t\equiv 0$. Under Assumption~\ref{ass:armax-regularity}, equality of the predictor
filters corresponds to equality of the regular \armaxtext{} representation. Thus, for the experiment to be informative enough, then two identical predictors, i.e., $\E[\Delta\hat{y}_t^2]=0$,  must imply that the systems are identical \cite{ljung1999system,gevers2008informative}, i.e., 
\begin{equation}\label{eq:informative}
\Delta W_u = 0 \text{ and } \Delta W_y = 0.
\end{equation}

We begin working out the difference between predictors in a closed-loop experiment. 
Substituting \eqref{eq:linearsystem} and \eqref{eq:linearcontroller} into \eqref{eq:prediction_diff} yields
\begin{equation}
\label{eq:predictordifference}
\begin{split}
\Delta \hat{y}_t
&=(\Delta W_u+\Delta W_y G)KSr_t\\
&\quad+(\Delta W_u+\Delta W_y G)Sd_t\\
&\quad+(\Delta W_y-\Delta W_uK)HSe_t, 
\end{split}
\end{equation}
where $S=(1+GK)^{-1}$ is the sensitivity function. By Assumption~\ref{ass:innovation-distribution}, $\ve_\ts$ has a positive
spectrum at all frequencies. Since we study the case $r_t\equiv 0$ and $d_t\equiv 0$, $\E[\Delta\hat{y}_t^2]=0$ implies
\begin{equation}\label{eq:informative2}
\Delta W_y=\Delta W_uK.
\end{equation}
Using \eqref{eq:deltaWs}, we can express \eqref{eq:informative2} as
\begin{equation}\label{eq:polyidentity}
  (A_2M+q^{-\orderdelay}B_2L)C_1=(A_1M+q^{-\orderdelay}B_1L)C_2.
\end{equation}

We now study conditions under which \eqref{eq:polyidentity} results in the desired situation defined by \eqref{eq:informative}. First, let $P_1$ denote the greatest common divisor of the closed-loop denominator polynomial $(A_1M+q^{-\orderdelay}B_1L)$ and $C_1$, viz.
\begin{equation}
\label{eq:gcd_poly}
    (A_1M+q^{-\orderdelay}B_1L)=P_1\bar{O}_1,\qquad C_1=P_1\bar{C}_1 .
\end{equation}
We normalize $P_1$ such that its leading coefficient is 1. Since $C_1$ has a leading coefficient of 1, then so has $\bar{C}_1$. The degree of $P_1$ is upper bounded by $n_p$ in  \eqref{eq:np_upper}. (If $(A_1M+q^{-\orderdelay}B_1L)$ and $C_1$ were coprime, then $\deg P_1=0$.) In a similar manner, we can construct $P_2$ for $(A_2M+q^{-\orderdelay}B_2L)$ and $C_2$. Then \eqref{eq:polyidentity} can be written as
\[
  P_2\bar{O}_2 P_1\bar{C}_1=P_1\bar{O}_1 P_2\bar{C}_2.
\]
Cancelling $P_1$ and $P_2$ from both sides, reduces it to
\[
\bar{O}_1\bar{C}_2=\bar{O}_2\bar{C}_1 .
\]
Hence $\bar{C}_1\mid \bar{O}_1\bar{C}_2$. Since, by construction, $\bar{C}_1$ and $\bar{O}_1$ are coprime, then $\bar{C}_1\mid\bar{C}_2$. Similary, from the same identity we have $\bar{C}_2\mid\bar{C}_1$. Thus $\bar{C}_1$ and $\bar{C}_2$ are mutually divisible and since both have leading coefficient 1, it follows that
\[
\bar{C}_1=\bar{C}_2,\qquad \bar{O}_1=\bar{O}_2 .
\]

Therefore, using \eqref{eq:gcd_poly}, we have
\[
\frac{A_1M+q^{-\orderdelay}B_1L}{P_1}
=
\frac{A_2M+q^{-\orderdelay}B_2L}{P_2},
\]
which can be re-arranged as
\[
\underbrace{(P_1A_2-P_2A_1)}_{\Delta A}M+q^{-\orderdelay} \underbrace{(P_1B_2-P_2B_1)}_{\Delta B}L=0 ,
\]
or, more compactly,
\begin{equation}
\label{eq:poly_compact}
 \Delta AM+q^{-\orderdelay}\Delta BL=0 .
\end{equation}
We can now express \eqref{eq:deltaWs} as
\begin{equation}
\label{eq:deltaWs_poly}
\begin{split}
\Delta W_u &=\frac{q^{-\orderdelay}B_1}{C_1} - \frac{q^{-\orderdelay}B_2}{C_2} =-\frac{q^{-\orderdelay}\bar{C}_1}{C_1C_2}\Delta B, \\
\Delta W_y &= \frac{A_2}{C_2}-\frac{A_1}{C_1}=\frac{\bar{C}_1}{C_1C_2}\Delta A.
\end{split}
\end{equation}

Since $L$ and $M$ are coprime by Assumption~\ref{ass:controller-coprime}, \eqref{eq:poly_compact} implies that there must be a polynomial $T$ such that
\[
\begin{cases}
\Delta A=q^{-\orderdelay}TL,\\
\Delta B=-TM,
\end{cases}
\]
and where the degree of $T$ bounded from above by
\[
\deg T\le \gamma\triangleq n_p+\min(n_a-n_\ell-\orderdelay,\;n_b-n_m).
\]

If $\gamma<0$, then the bound $\deg T\le \gamma$ cannot be satisfied by any nonzero polynomial, since every nonzero polynomial must have non-negative degree. Then $T\equiv0$ and, therefore, \eqref{eq:deltaWs_poly} reduces into \eqref{eq:informative}.
\end{pf}

\begin{remark}
The proof follows the same general line as in \cite{gevers2008informative} but generalizes the result therein by relaxing the requirement for coprimeness between $A(q)M(q)+q^{-\orderdelay}B(q)L(q)$ and $C(q)$. The new result is, moreover, adapted to account for the explicit extra delay $q^{-\orderdelay}$.
\end{remark}

From \eqref{eq:lambdacondition}, we see that when the controller order and/or delay is sufficiently high relative to the orders of  the system, then no probing signal is required for identifiability. On the other hand, we now show that sufficiently rich probing signal $d_t$ ensures identifiability, no matter the value of $\gamma$.

\begin{definition}[Persistent excitation \cite{soderstrom1989system}]
A quasi-stationary probing signal $d_t$ is said to be persistently exciting of order $\orderexcit$ if
\begin{equation}
\E\left[\begin{bmatrix}
d_{t-1}\\
d_{t-2}\\
\vdots\\
d_{\ts-\orderexcit}
\end{bmatrix} \begin{bmatrix}
d_{t-1}\\
d_{t-2}\\
\vdots\\
d_{\ts-\orderexcit}
\end{bmatrix}^\top \right] \succ0.
\label{eq:pe_definition}
\end{equation}
\end{definition}

\begin{theorem}\label{thm:withd}
Consider a closed-loop experiment with constant reference in which $\gamma \geq 0$ (see \eqref{eq:lambdacondition}). Suppose $d_t$ is a bounded discrete-valued probing signal\footnote{The design $\vd_{\ts}$ we consider below will also be discrete-valued at any $\ts$.} that is persistently exciting of order 
\[
\orderexcit \ge \gamma .
\]
Then $\{ (\vpinput_\ts,\voutput_{\ts} ) \}$ is informative enough to identify the \armaxtext{} model.
\end{theorem}
\begin{pf}
Following the reasoning in the proof of Theorem~\ref{thm:ealone}, we study the implications of $\E[\Delta\hat{y}_t^2]=0$. Using \eqref{eq:predictordifference}, it is equivalent to
\begin{equation}\label{eq:withcross}
    \E[ \big( (\Delta W_u + \Delta W_yG)Sd_t + (\Delta W_y-\Delta W_u K)HSe_t \big)^2] = 0.
\end{equation}

We first consider the case when $\Delta W_y = \Delta W_u K$. Then  \eqref{eq:withcross} reduces to \(
\E[(\Delta W_u d_t)^2] = 0.
\)

Since \eqref{eq:deltaWs_poly} holds in this case, we obtain 
\[
\E\!\left[\left(\frac{q^{-\orderdelay}\bar{C}_1 TM}{C_1C_2} d_t\right)^2\right] = 0.
\]
Because $\vd_\ts$ is persistently exciting of $\orderexcit \ge \gamma$, this implies $T\equiv 0$ so that \eqref{eq:informative} follows.

Next, consider the complementary case $\Delta W_y \neq \Delta W_u K$. Suppose there exists some $\Delta W_u \neq 0$ and/or $\Delta W_y\neq 0$ that satisfies  \eqref{eq:withcross}. Then we have
\[
\underbrace{(\Delta W_u + \Delta W_yG)S}_{X(q)}d_t = \underbrace{-(\Delta W_y-\Delta W_u K)HS}_{Z(q)}e_t
\quad \text{(a.s.)}
\]
We now show that this cannot hold.
If $X(q)=0$, we have
\[0 = Z(q)e_\ts
\quad \text{(a.s.)}\]
Since \(Z(q)\neq 0\) by assumption, the RHS is the output of a nonzero linear filter driven by the white process \(e_t\) with a continuous distribution over the real line, and therefore cannot vanish almost surely. On the other hand, if $X(q)\neq 0$, then we have
\[
d_t = \frac{Z(q)}{X(q)} e_\ts
\quad \text{(a.s.)}
\]
The RHS, by Assumption~\ref{ass:innovation-distribution}, has a continuous distribution. Since \(\vd_\ts\) is bounded and discrete-valued, the above equality cannot hold almost surely.
Hence there exists no nontrivial $\Delta W_u$ and $\Delta W_y$ that satisfy \eqref{eq:withcross}, thereby implying \eqref{eq:informative} also in the complementary case.
\end{pf}
Theorem~\ref{thm:withd} provide general conditions on the probing signal  $d_t$  that  \emph{ensure} informativeness. Below we will recursively design such signals so as to reduce the estimation errors of $\para$.

\subsection{Feasibility of perturbation limits}

We now analyze the user-specified perturbation limits $(\ydmin, \ydmax)$ in
\eqref{eq:optim_ideal}. 

\begin{definition}[Recursive feasibility]
The specified perturbation limits $(\ydmin, \ydmax)$ are said to be recursively feasible if satisfying $\ydmin \leq \yd_{\ts} \leq \ydmax$ implies that
$$\ydmin \leq \yd_{\ts+1} \leq \ydmax$$
is feasible.
\end{definition}

Thus when the chosen limits are recursively feasible, the sequence of constraints in \eqref{eq:optim_ideal} can be relaxed to a single step.

The perturbation is shaped by the unknown (load) sensitivity function $G_d$ \cite[ch.~11]{astrom2021feedback}:
\begin{equation}
\yd_{\ts} = \underbrace{\frac{G(q;\para)}{1+ G(q;\para) K(q)}}_{G_d(q; \para)} \vd_{\ts} = \sum^{\infty}_{i=1} \widetilde{g}_i \vd_{\ts-i}.\label{eq:sensitivity}
\end{equation}
Since the closed-loop system is stable,  $G_d(q; \para)$ is  also stable. We study the perturbation using the impulse response $\{  \widetilde{g}_i\}$ of the sensitivity function:
\begin{equation}
    G_d(q;\para) = \frac{q^{-\orderdelay}B(q;\para)M(q)}{A(q;\para)M(q)+q^{-\orderdelay}B(q;\para)L(q)} \triangleq \frac{\widetilde{B}(q;\para)}{\widetilde{A}(q;\para)}\label{eq:impulse1}
\end{equation}
where the coefficients for the polynomials $\widetilde{A}$ and $\widetilde{B}$ and  are given by convolutions:
\begin{equation}
\widetilde{b}_i = \begin{cases}
0, & i=0,\ldots,\orderdelay,\\[1mm]
\displaystyle \sum_{j=0}^{i-\orderdelay} b_j m_{i-\orderdelay-j},
& i=\orderdelay+1,\ldots,\orderdelay+n_b+n_m .
\end{cases},
\end{equation}
and \begin{equation}
   \tilde a_i=
\sum_{j=0}^{i} a_j m_{i-j}
+
\sum_{j=0}^{i-\orderdelay} b_j \ell_{i-\orderdelay-j},
\end{equation}
omitting $\para$ for notational convenience. These are readily computable given the parameter vector in \eqref{armax}. To obtain the impulse response in \eqref{eq:sensitivity}, we use the relation 
\(\widetilde{A}(q) G_d(q) = \widetilde{B}(q). \) By applying the operators on both sides, we obtain the relation: 
\begin{equation*}
    \begin{split}
        \yd_{\ts} &= \sum_{i=1}^{n_b+n_m}\widetilde{b}_i\vd_{\ts-i} -\sum_{i=1}^{\max(n_a+n_m,n_b+n_\ell)}\widetilde{a}_i\yd_{\ts-i}\\&= \sum_{i=1}^{n_b+n_m}\widetilde{b}_i\vd_{\ts-i} -\sum_{i=1}^{\max(n_a+n_m,n_b+n_\ell)}\widetilde{a}_i \Big(\sum_{j=1}^{\infty}\widetilde{g}_jd_{\ts-i-j} \Big). 
    \end{split}
\end{equation*} 
 Comparing with \eqref{eq:sensitivity}, this gives a recursive relation for the impulse response:
\begin{equation}
\widetilde{g}_i = \widetilde{b}_i - \sum_{j=1}^i \widetilde{a}_j \widetilde{g}_{i-j},\label{eq:impulse_coef}
\end{equation}  where $\widetilde{g}_0 = \widetilde{b}_0 = 0$.

Suppose the probing experiment starts at time $\ts-\K+1$, where $k>\orderdelay$, then the output perturbation equals
\begin{equation}
    \yd_{\ts+1} = \sum_{i=1}^\K \widetilde{g}_i\vd_{\ts+1-i}
\label{eq:sensitivityimpulse}
\end{equation}
where, due to the extra delay, \(\tilde g_i=0\) for \(i=1,\ldots,\orderdelay\). Hence, the impact of $d_\ts$ can not be observed until time  $\ts+\orderdelay+1$:
\begin{equation}\label{eq:finitewindowperturb}
\delta_{\ts+\orderdelay+1}
=
\sum_{i=1}^{k-\orderdelay}\tilde g_{\orderdelay+i} d_{t+1-i} = \widetilde{g}_{\orderdelay+1} \vd_{\ts} + \constb_{\ts}.  \vspace{-0.5em}
\end{equation}
where 
\begin{equation}
 \constb_\ts \triangleq \begin{bmatrix} \widetilde{g}_{\orderdelay+2} & \widetilde{g}_{\orderdelay+3} & \cdots & \widetilde{g}_{\K} \end{bmatrix} \begin{bmatrix}
\vd_{\ts-1} \\ \vd_{\ts-2} \\ \vdots \\ \vd_{\ts-\K +1 + \orderdelay}
\end{bmatrix}
\label{eq:h_history} \vspace{-0.5em}
\end{equation}
is fixed at time $\ts$. 

Using \eqref{eq:finitewindowperturb}, we now study the recursive feasibilty of the perturbation limits. For clarity, we restrict the analysis to the symmetric  limits on the probing signal:
\[
d_t\in[-\dmax,\dmax],
\]
which is also the setting used in our numerical experiments. It is also helpful to define the following vectors
\begin{equation*}
\begin{split}
g_\K &\triangleq \begin{bmatrix} \widetilde{g}_{\orderdelay+1} & \widetilde{g}_{\orderdelay+2} & \cdots & \widetilde{g}_{\K-1} &\widetilde{g}_{\K} \end{bmatrix}^\top \\
s_\K &\triangleq \begin{bmatrix} \widetilde{g}_{\orderdelay+2} & \widetilde{g}_{\orderdelay+3} & \cdots & \widetilde{g}_{\K} &0\end{bmatrix}^\top.
\end{split}
\end{equation*}

\begin{proposition}\label{prop:recursive_feasibility}
If the specified perturbation limits satisfy
\begin{equation}
\label{eq:feasible_limits}
\begin{split}
\ydmin \leq -\vd_{\max}\nu(\K,\para) \quad \text{and} \quad
\ydmax \geq \vd_{\max}\nu(\K,\para),
\end{split}
\end{equation}
where 
\begin{equation}
    \nu(\K,\para)= \left( \inf_{v\in[0,1)} \frac{\|s_k-v g_k\|_1-|\constA|}{1-v} \right)_+,
\label{eq:extremal_nu}
\end{equation}
then the limits are recursively feasible.
\end{proposition}
The proof is given in Appendix~\ref{ap:proof-recursive-feasiblity}.

Thus when the specified limits satisfy \eqref{eq:feasible_limits}, the  design problem \eqref{eq:optim_ideal} with multi-step constraints can be relaxed into a problem with a one-step constraint:
\begin{equation}
    \begin{split}
         \min_{\vd_\ts} & \: J_{\ts+1} \\
         \text{s.t. } & \: \dmin \leq \vd_\ts \leq \dmax\\
         & \: \ydmin - \constb_\ts(\para) \leq  \widetilde{g}_{\orderdelay+1}(\para)\vd_{\ts}\leq \ydmax - \constb_\ts(\para),\\
         \end{split}
         \label{eq:relaxed}
\end{equation}
using \eqref{eq:finitewindowperturb}. When the  limits do not satisfy \eqref{eq:feasible_limits}, it is possible that for certain $t$, \eqref{eq:relaxed} becomes infeasible. For those instances, the default is to set $d_t=0$ since feasibility is then \emph{recovered} in finite time by exponential closed-loop stability.




\section{Recursive Design Method}
\label{sec:method}

The recursive design of the probing signal is based on the recursive prediction error method \rpemtext{} for \armaxtext{} systems \cite{soderstrom1989system,ljung1999system}. Here we first discuss a parameterization that is appropriate for recursive estimation when the system delay is unknown. We then present a suitable design criterion and, finally, a closed-form solution of the designed probing signal.

\subsection{Parameterization}

Since the additional input delay $\orderdelay$ is unknown a priori, the original parameterization $\para$ is not readily suitable for recursive estimation. 
Here we use the expanded parameter vector  \[\paraexp = [\beta_1\;\dots\;\beta_{n_\beta}\;a_1\;\dots\;a_{n_a}\;c_1 \dots\; c_{n_c}]^{\top}\] 
instead, where the the associated input polynomial is
\[\overline{B}(q;\paraexp) = \sum_{i=1}^{n_\beta}\beta_iq^{-i} \quad \text{and} \quad n_\beta = n_b+\orderdelay^{\max}.\]
This expanded parameterization is sufficient to cover the original input dynamics together with all admissible values of the unknown extra delay $\orderdelay^{\max}$, which is implicitly encoded by the location of the first nonzero input coefficient.

\subsection{Design criteria}
Under the prediction error framework, the large-sample error covariance matrix of $\paraexpest_{\ts+1}$ is given by
$$P_{\ts+1} = \frac{\lambda}{\ts+1} \big(\mathbb{E}[\grad_{\ts+1} \grad_{\ts+1}^\top]\big)^{-1},$$
where $\grad_{\ts+1}$ is the (negative) gradient of the one-step-ahead prediction error \cite[ch~7.]{soderstrom1989system}:
\begin{equation}
\begin{split}
\grad_{\ts+1} &\equiv -\frac{\partial\varepsilon_{\ts+1}}{\partial\paraexp}\\ &= -\frac{\partial}{\partial\paraexp} \left[ \frac{A(q; \paraexp)}{C(q; \paraexp)}\voutput_{\ts+1} - \frac{\overline{B}(q; \paraexp)}{C(q; \paraexp)}\vpinput_{\ts+1}  \right]. \label{eq:gradient}
\end{split}
\end{equation}
We can exploit this structure to compute an estimated error covariance matrix that forms the basis of the design criterion  $J_{\ts+1}$ in \eqref{eq:relaxed}. 

Since the element-wise derivatives are obtained by filtered signals
\begin{equation*}
\begin{split}
\frac{\partial \varepsilon_{\ts+1}}{\partial \beta_i} &= -\frac{1}{C(q;\paraexp)}\vpinput_{\ts-i+1}, \\
\frac{\partial \varepsilon_{\ts+1}}{\partial a_i} &= \frac{1}{C(q;\paraexp)}\voutput_{\ts-i+1} \\
\frac{\partial \varepsilon_{\ts+1}}{\partial c_i} &= -\frac{1}{C(q;\paraexp)}\varepsilon_{\ts-i+1} ,
\end{split}
\end{equation*}
the negative gradient in \eqref{eq:gradient} can be computed recursively as:
\begin{equation}
\grad_{\ts+1} = \regressor_{\ts+1}-c_1\grad_{\ts}-\dots-c_{n_c}\grad_{\ts-n_c+1},
\label{eq:score-plugin}
\end{equation}
where we define the vector
\begin{equation*}
\begin{split}
\regressor_{\ts+1} &= [\vpinput_\ts \: \cdots \: \vpinput_{\ts-n_\beta+1}\: -\voutput_\ts \: \cdots \:-\voutput_{\ts-n_a+1} \\
&\qquad \varepsilon_{\ts} \: \cdots \: \varepsilon_{\ts-n_c+1}]^{\top},
\end{split}
\end{equation*}
see also \cite[ch~10.3]{ljung1999system}.  Let $\gradest_{\ts+1}$ denote the negative gradient evaluated at $\paraest_\ts$ and let $\widehat{\varepsilon}_{\ts}$ be the estimated prediction error. Then using relation \eqref{eq:score-plugin} with estimates enables the re-use of previous estimates and thus an efficient online evaluation.  The unknown covariance matrix is then estimated as
\begin{equation}
    \widehat{P}_{\ts+1} = \noisevest \Big(\sum_{i=1}^{\ts+1}\gradest_i\gradest_i^{\top} \Big)^{-1} = \noisevest R_{\ts+1} 
\label{eq:estimatedcovariance}
\end{equation}
where $\noisevest$ denotes the estimated variance of $e_t$ and $R_{\ts+1} \equiv (R^{-1}_{\ts} + \gradest_{\ts+1}\gradest_{\ts+1}^\top)^{-1}$ can be computed recursively using the matrix inverse lemma. (The recursion may be initialized using  $R_0 \propto I$.)  We note that only $\gradest_{\ts+1}$ is dependent on the probing signal $\vd_{\ts}$ to be designed.

Given \eqref{eq:estimatedcovariance}, we consider two recursively computable design criteria in \eqref{eq:relaxed}:
\begin{equation}
J_{\ts+1} = |\widehat{P}_{\ts+1}| \quad \text{or} \quad J_{\ts+1} = \text{tr}\left\{ \widehat{P}^{-1}_{\ts} \widehat{P}_{\ts+1} \right\},
\label{eq:designcriteria}
\end{equation}
which are both scale invariant with respect to the parameter vector. The determinant criterion (aka. D-criterion) is a common choice, whereas the weighted trace criterion (aka. L-criterion \cite{federov1972optimalexp}) provides a measure of successive gain in an online experiment.

\subsection{Recursive closed-form probing signal design}

The D-criterion in \eqref{eq:designcriteria} can be expressed as
\begin{equation*}
J_{\ts+1} = \left(1+\gradest_{\ts+1}^\top R_{\ts}\gradest_{\ts+1}\right)^{-1}|R_{\ts}|
\end{equation*}
using the matrix determinant lemma, while the L-criterion in \eqref{eq:designcriteria} can be expressed as
\begin{equation*}
J_{\ts+1} = \left(1+\gradest_{\ts+1}^\top R_{\ts}\gradest_{\ts+1}\right)^{-1} + \text{constant.}
\end{equation*}
Thus, minimization of either criterion is equivalent to maximizing $\gradest_{\ts+1}^\top R_{\ts}\gradest_{\ts+1}$. We can now finally express \eqref{eq:relaxed} as the follwing \emph{adaptive design problem}:
\begin{equation}
    \begin{split}
          \max_{\vd_\ts} &  \;\gradest_{\ts+1}^\top R_{\ts}\gradest_{\ts+1}\\
         \text{s.t. } & \dmin \leq \vd_\ts \leq \dmax\\
         &    \ydmin - \constb_\ts \leq \constA \vd_{\ts}\leq \ydmax - \constb_\ts.
         \end{split}
\label{eq:designproblem}
\end{equation}

To arrive at a closed-form solution of \eqref{eq:designproblem}, we partition $R_{\ts}$ and
$\gradest_{\ts+1}$ as:

\begin{equation}\label{eq:rmatrix}
R_{\ts} =\begin{bmatrix}
    R_{11}&R_{12}\\
    R_{12}^\top&R_{22}
\end{bmatrix},\quad
        \gradest_{\ts+1} = \begin{bmatrix}
            \tau_{\ts+1}\\
            \xi_{\ts+1}
        \end{bmatrix},
\end{equation}
where
$R_{11}>0$ and $\tau_{\ts+1}$ are scalars.

\begin{theorem}\label{thm:closed-form-solution}
The adaptive probing signal that solves \eqref{eq:designproblem} is given by:
\begin{equation}
\boxed{
    \vd_\ts^* = \begin{cases}
      \vd_l & \text{if $-\frac{R_{12}\xi_{\ts+1}}{R_{11}} - \widehat{\vinput}_\ts> \frac{\vd_l+\vd_u}{2}$}\\
      \vd_u & \text{otherwise}    \end{cases}\label{eq:solution-siso}
      }
\end{equation} where $$\widehat{\vinput}_\ts \equiv \vinput_{\ts} - \widehat{c}_1\tau_{\ts}-\dots-\widehat{c}_{n_c}\tau_{\ts-n_c+1},$$
\begin{equation}\label{eq:dl}
\vd_l = \max\left( \dmin,\ \min\left( \frac{\ydmin - \constb_\ts}{\constA},\ \frac{\ydmax - \constb_\ts}{\constA} \right) \right)\end{equation}
and
\begin{equation}\label{eq:du}
\vd_u=
\min\left( \dmax,\ \max\left( \frac{\ydmin - \constb_\ts}{\constA},\ \frac{\ydmax - \constb_\ts}{\constA} \right) \right).
\end{equation}
The quantities in \eqref{eq:solution-siso} are evaluated using the current estimate $\paraexpest_{\ts}$. 
\end{theorem}

\begin{pf}
Using \eqref{eq:score-plugin},  we have that 
\begin{equation}
        \gradest_{\ts+1} = \begin{bmatrix}
            \tau_{\ts+1}\\
            \xi_{\ts+1}
        \end{bmatrix}\,
        = \begin{bmatrix}
            \underbrace{\vinput_{\ts} - \widehat{c}_1\tau_{\ts}-\dots-\widehat{c}_{n_c}\tau_{\ts-n_c+1}}_{\widehat{\vinput}_{\ts}}\\
            \xi_{\ts+1} 
        \end{bmatrix} + \begin{bmatrix}
                \vd_{\ts} \\
                0
            \end{bmatrix}.
    \label{eq:grad_split}
\end{equation}

 Substituting \eqref{eq:grad_split} into the quadratic function \eqref{eq:designproblem} yields:
\begin{equation}
    \begin{split} 
         \max_{\vd_\ts} & \; R_{11}\vd_{\ts}^2+2(R_{11}\widehat{\vinput}_\ts+R_{12}\xi_{\ts+1})\vd_{\ts} \\
         \text{subject to} & \; \vd_l \leq \vd_\ts\leq \vd_u.
    \end{split}
\end{equation}
The quadratic function is convex and symmetric with respect to the point:
$$\vd_m = -\frac{R_{12}\xi_{\ts+1}}{R_{11}} - \widehat{\vinput}_\ts.$$
Therefore its maximum is located at one of the boundaries of the feasible interval  
$[\vd_l,\vd_u]$. By comparing
$\vd_m$ with the midpoint 
$\frac{\vd_l+\vd_u}{2}$, we can conclude
\begin{itemize}
    \item If $\vd_m>\frac{\vd_l+\vd_u}{2}$, the optimal value is attained at $\vd_l$.
    \item  If $\vd_m<\frac{\vd_l+\vd_u}{2}$, the optimal value is attained at $\vd_u$.
\end{itemize}
This proves the theorem.
\end{pf}

\begin{remark}
The  expressions of $\vd_l$ and $\vd_u$ take into account the possibility of $\constA$ being negative. Note that they depend on the unknown extra delay $\orderdelay$, via the second constraint in \eqref{eq:designproblem}. In practice, this can be estimated from $\paraexpest_{\ts}$. A pragmatic estimator of \(\widehat \orderdelay\) is
\begin{equation}\label{eq:estimate_delay}
\widehat \orderdelay
=
\max\left\{
\rho\in\{1,\dots,\orderdelay^{\max}\} :
\frac{\max_{1\le i\le \rho}|\widehat{\beta}_i|}{|\widehat{\beta}_{\rho+1}|}\le \varrho
\right\},
\end{equation}
where $\varrho \ll 1$ is a chosen threshold and, by default, \(\widehat \orderdelay = 0\) if no solution exists. This gives an explicit parametric
time-delay estimation method, see also~\cite{bjorklund2003review}.
\end{remark}

It follows from \eqref{eq:solution-siso} that $\vd^*_{\ts}$ is bounded and discrete-valued, which conforms to the premise in Theorem~\ref{thm:withd}. For completeness, the resulting recursive procedure is summarized in
Appendix~\ref{ap:alg1} Algorithm~\ref{alg:sequential-probing}.

A complete characterization of this  probing signal is nontrivial as it is in part determined by past information described by the sigma-field
\[
\mathcal F_{t-1}
=
\sigma\!\left(
\{u_i,\widetilde{y}_i,r_i,e_i,d_i,\paraexpest_i,\widehat{P}_i\}_{i\le t-1}
\right).
\]
However, we will now show that conditional on $\mathcal{F}_{t-1}$, the binary variable  $\vd^*_{\ts}$ is random. Then we show that the probing signal is persistently exciting.

Due to linearity of the closed-loop system, we can express the left-hand side of the inequality in \eqref{eq:solution-siso} in terms of the exogenous random noise $e_t$. More specifically, conditional on $\mathcal F_{t-1}$, we have
$$-\frac{R_{12}\xi_{\ts+1}}{R_{11}} - \widehat{\vinput}_\ts = f_{1,t}e_t+f_{0,t},$$
where the coefficients $f_{0,t}$ and $f_{1,t}$ are given. Since the regressors are bounded, then so are these coefficients and $f_{1,t} \neq 0$, apart from a degenerate tuning of $\ell_0$ in \eqref{eq:controller_polynomial}.\footnote{It can be shown that
\[
f_{1,t}
=
\left[ -\sum_{i=1}^t \tau_i\xi_i^\top(\xi_i\xi_i^\top)^{-1} \right] v_{n_\beta }^\top
+
\ell_0,
\]
where $v_{n_\beta}$ is a standard basis vector \(\ell_0\) is given by the controller since \(e_t\) enters $u_\ts$ through \eqref{eq:linearcontroller}.}
Thus the conditional probability of either outcome in \eqref{eq:solution-siso} is determined by
\[
\mathbb{P}\left(f_{1,t}e_t+f_{0,t}
> \frac{d_{l}+d_{u}}{2} \: \big| \: \mathcal{F}_{t-1} \right)
\]
and, together with the fact that $d_{l}$ and $d_{u}$ are bounded, the probability is bounded away from 0.

\begin{proposition}
If $\mathbb{P}\!\left(d^*_t\mid\mathcal F_{t-1}\right)$ is bounded away from zero, then \eqref{eq:pe_definition} holds for any $\orderexcit$. That is, $\vd^*_{\ts}$ is persistently exciting of any arbitrary order and the resulting data $\{ (\vpinput_\ts,\voutput_{\ts} ) \}$ is informative enough to identify the \armaxtext{} system.
\end{proposition}

\begin{pf}
We drop the asterisk in $d^*_{\ts}$ for notational simplicity. Since $\vd_{\ts}$ is binary, $|d_{u,t}-d_{l,t}|\ge\eta > 0$. Let $p = \mathbb{P}(d_t =d_{u,t} \mid\mathcal F_{t-1})$, therefore
\[
\mathbb{V}(d_t\mid\mathcal F_{t-1})
\ge
p(1-p)\eta^2>0.
\] 

Next, consider the inner product $x^\top D_t$, where
\[
D_t=[d_t,d_{t-1},\ldots,d_{t-\orderexcit+1}]^\top
\] and an arbitrary nonzero vector 
\[x=[x_0,\dots,x_{\orderexcit-1}]^\top \neq 0.\]
Now chose the smallest $j$ for which the coefficient \(x_j\) is nonzero and express the inner product as:
\[
x^\top D_t
=
x_j d_{t-j}
+
\sum_{i=j+1}^{\orderexcit-1} x_i d_{t-i}.
\]
Then
\begin{equation*}
\begin{split}
\mathbb{V}\!\left(x^\top D_t \mid \mathcal F_{t-j-1}\right) &= x_j^2
\mathbb{V}\!\left(d_{t-j}\mid \mathcal F_{t-j-1}\right) \\
&\ge x_j^2 p(1-p)\eta^2 >0 .
\end{split}
\end{equation*}
By marginalization over $\mathcal F_{t-j-1}$, we have
\begin{equation*}
    \begin{split}
     \mathbb E\!\left[(x^\top D_t)^2\right] &=\mathbb E\!\left[\mathbb E\!\left[(x^\top D_t)^2\mid \mathcal F_{t-j-1}\right] \right]   \\
     &=\mathbb E\!\left[
\mathbb{V}\!\left(x^\top D_t \mid \mathcal F_{t-j-1}\right)
\right]\\
& \quad+
\mathbb E\!\left[
\left(
\mathbb E\!\left[x^\top D_t \mid \mathcal F_{t-j-1}\right]
\right)^2
\right]\\
&\ge x_j^2 p (1-p)\eta^2>0.
    \end{split}
\end{equation*}
Thus for any nonzero $x$, we have
\[\mathbb E\!\left[(x^\top D_t)^2\right] = x^\top \mathbb E[D_tD_t^\top]x>0 \]
and therefore \eqref{eq:pe_definition} holds for any $\orderexcit$. Thus the discrete-valued probing signal satisfies the condition in Theorem~\ref{thm:withd}.
\end{pf}

\begin{remark}
Since the closed-loop system is assumed to be exponentially stable, the past samples $\vd_{\ts-\K+1}$ have a negligible impact on the perturbation \eqref{eq:sensitivityimpulse} for large $\K$. In a practical implementation of \eqref{eq:solution-siso}, one can therefore use a large but fixed horizon $\K$ and store only the last $\K$ samples of $\vd_{\ts}$. The boundaries in \eqref{eq:solution-siso} are determined by $\constA(\para)$ and $\constb_\ts(\para)$ which are re-evaluated using the recursive estimates $\paraest_{\ts-1}$. This leads to an adaptation of the estimated constraints that steers the system towards the specified perturbation limits (as illustrated below).
\end{remark}

\section{Numerical experiments}\label{sec:numeric}

We evaluate the designed probing signal in terms of the resulting estimation errors using \rpemtext{} with input-output data $\{ (\vpinput_\ts,\voutput_{\ts} ) \}$ as well as how well it steers the output perturbations within specified limits. The designed signal is compared against two baseline $\vd_{\ts}$: one where it is zero and the other is a pseudorandom binary signal (\prbstext{}), which provides an informative probing for linear systems but does take into account the resulting perturbation of the output. Finally, we turn to study the frequency characteristics of the designed probing signal.

\subsection{Experimental settings}

For all the probing signals, we consider symmetric constraints, i.e., $\dmax=-\dmin$. The proposed design also uses symmetric perturbation limits $\ydmax=-\ydmin$ and a fixed horizon of $k=50$ samples. 

In the systems considered below, we use a PI-controller with constant reference signal $r_{\ts} \equiv 1$. This provides a means of assessing the magnitude of the perturbation limits. For instance, $\ydmax=0.10$ means that the user tolerates up to 10\% output perturbations relative to the reference signal. When ground truth is known, one can also compare this against the standard deviation of the output disturbance $H(q; \para) e_t$, which is
\[
    \sigma_v
    =
    \sqrt{\lambda}\left(
        \frac{1}{2\pi}
        \int_{-\pi}^{\pi}
        \left|H(e^{\mathrm{i}\omega};\para)\right|^2
        \,d\omega
    \right)^{1/2}.
\]
where we use $\sqrt{\lambda} = 0.01$ throughout.
The nominal sampling period is \(T_s=0.01\) seconds. 

The \rpemtext{} \cite{MATLABRPEM} is employed for recursive estimation of $\paraexp$, using a time-varying forgetting factor given by $1-0.02\cdot(0.998)^\ts$ \cite{ljung1999system}. An initial estimate is formed using the first 200 samples during which there is no probing signal. For the delay estimator, we set $n^{\max}_k = 3$.



\subsection{Performance analysis}

We first consider the following system (\armaxtext{}-1):
\begin{equation}
    \begin{split}
        A(q) &= 1 - 0.9062 q^{-1} + 0.4344 q^{-2} - 0.1829 q^{-3} \\  
        B(q) &= 0.57 q^{-1} - 0.38 q^{-2} + 0.118 q^{-3}\\
        C(q) &= 1 + 0.2 q^{-1}.
    \end{split}\label{armaxcase1}
\end{equation}
with no additional input delay $(\orderdelay=0)$. In this case the output disturbance has a standard deviation $\sigma_v \approx 0.011$. The system is regulated by a PI-controller
\begin{equation}
    K(q) = \frac{0.005607+0.005607 q^{-1}}{1 - q^{-1}}.\vspace{-0.5em}
\end{equation}
For this system \(\gamma = 3\), which by Theorem~\ref{thm:withd} means that a probing signal \(\vd_\ts\) with persistent excitation of order \(\orderexcit\ge 3\) is sufficient for identifiability.  

\begin{figure}[!th]
    \centering
    \includegraphics[width=0.82\linewidth]{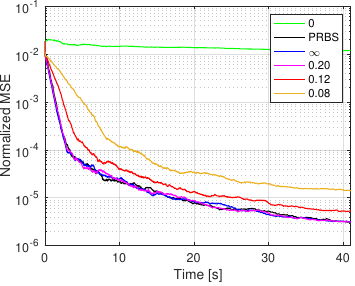}
    \caption{Mean squared error $\text{MSE}_\ts$ \eqref{eq:MSEparameter} for \armaxtext{}-1 when using \prbstext{} and designed probing signals with varying $\ydmax$. The signal are compared with their resulting perturbations in Figure~\ref{fig:perturbationmagnitude}.}
    \label{fig:armax-pi-mse-para}
    \vspace{-0.5em}
\end{figure}

We consider $\dmax=0.3$ and vary $\ydmax$. Figure~\ref{fig:armax-pi-mse-para} confirms that larger perturbation limits  \(\ydmax\) increases the convergence rate  of the parameter estimates, as evaluated by
\begin{equation}
\text{MSE}_\ts = \frac{\mathbb{E}[\|\paraexp - \paraexpest_{\ts}\|^2]}{\|\paraexp\|^2},
\label{eq:MSEparameter}
\end{equation}
using 100 Monte-Carlo runs.
Using the ground truth, we evaluate \(\nu=0\) in  \eqref{eq:extremal_nu} and conclude that any limit $\ydmax>0$ is recursively feasible. We also see that the zero probing baseline \(\vd_\ts=0\) yields a very slow parameter convergence, if at all, thereby demonstrating the insufficient excitation from \(e_\ts\) alone. At the other extreme, the \prbstext{} baseline  essentially provides a performance limit for identification.

\begin{figure}[!th]
    \centering
    \includegraphics[width=0.82\linewidth]{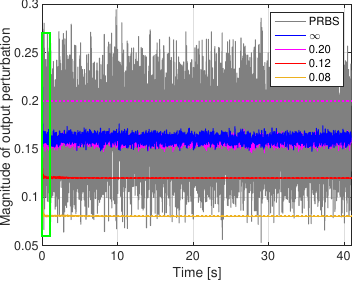}
    \caption{Magnitude of  perturbation $\mathbb{E}[|\yd_{\ts}|]$ for \armaxtext{}-1 when using \prbstext{} and the designed probing signals $\vd_{\ts}$ for various constraints 
$\ydmax$. The magnitude can be compared to the reference signal $r_{\ts} \equiv 1$. Dotted lines show user-specified limits $\ydmax$. 
}\label{fig:perturbationmagnitude}
 \vspace{-0.5em}
\end{figure}

\begin{figure*}[!t]
    \centering
\subfloat{\includegraphics[width=0.41\linewidth]{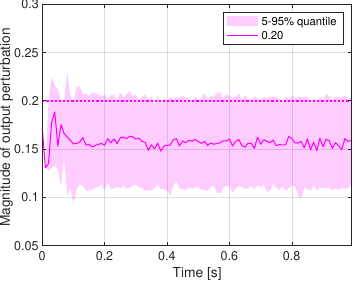}\label{fig:perturbationmagnitudefirst020}}\subfloat{\includegraphics[width=0.41\linewidth]{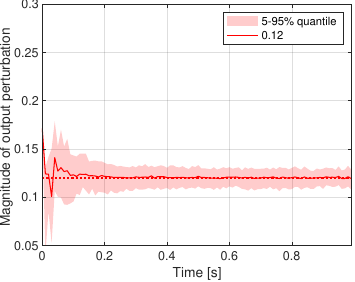}\label{fig:perturbationmagnitudefirst012}}
    \caption{Zoomed-in view (first 100 samples) of the perturbation magnitude corresponding to the green box in  Fig.~\ref{fig:perturbationmagnitude}. The solid line represents the mean over 100 Monte Carlo runs, and the shaded region shows the 5--95\% quantile band. (a): \(\ydmax=0.20\). (b): \(\ydmax=0.12\).}  \vspace{-0.5em}
\end{figure*}

The identification performance is of course to be traded off against the resulting perturbations $\yd_{\ts}$. The magnitude of the perturbations under different specified limits is illustrated in Figures~\ref{fig:perturbationmagnitude},~\ref{fig:perturbationmagnitudefirst020}, and~\ref{fig:perturbationmagnitudefirst012}. These include zoomed-in views and \(5\)--\(95\%\) quantile bands over 100 Monte Carlo runs. As expected the \prbstext{} baseline produces substantially larger perturbations,
often exceeding \(25\%\) of the reference signal, as compared to the adaptively designed signal even when there is no limit $\ydmax = \infty$. To explain this, recall that the constraints in \eqref{eq:designproblem} imposed by the limits are adaptively refined by $\paraexpest_{\ts-1}$. Thus the perturbations can exceed limits especially for the initial samples, but as $\paraexpest_{\ts-1}$ improves the adaptive design rapidly brings them under control.

The figures also show when the perturbation limits become active. For \(\ydmax=0.20\), the design is rather limited by $\dmax$ in which case perturbation magnitudes are very similar to the unconstrained case \(\ydmax=\infty\). For the tighter limit \(\ydmax=0.12\), the perturbation constraints on the design becomes more frequently active and shows larger initial violations, due to the combined effect of lower parameter accuracy during the transient and the stricter admissible perturbation level.

\begin{figure}[!th]
     \centering   \includegraphics[width=0.82\linewidth]{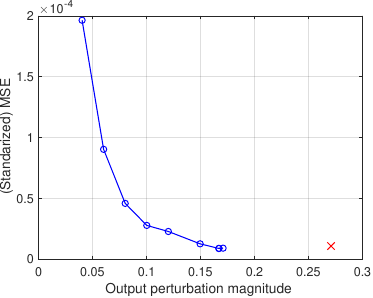}
     \caption{Model error $\overline{\text{MSE}}$ \eqref{eq:MSEmodel} versus the peak value of $\mathbb{E}[|\yd_{\ts}|]$ for \armaxtext{}-1, evaluated over the final second of the experiment. The perturbation magnitude can be compared to the reference signal $r_{\ts} \equiv 1$. The circles display the performance of the constrained designs, sorted with respect to $\ydmax \in \{0.04, 0.06, 0.08, 0.10, 0.12, 0.16, 0.20, \infty \}$. The red cross denotes \prbstext{}, for comparison.}
     \label{fig:tradeoff} \vspace{-0.5em}
 \end{figure}

To further illustrate the trade off between experimental performance and cost, we consider the accuracy of the estimates in terms of the resulting frequency response $G(e^{i\omega}; \paraest_\ts)$, which is of particular interest when seeking an improved  controller. The model error is quantified here by the standardized mean squared error measure:
\begin{equation}
   \overline{\text{MSE}} =  \frac{ \int \mathbb{E}[| G(e^{i\omega}) - G(e^{i\omega}; \paraest_\ts) |^2] d\omega}{\int |G(e^{i\omega})|^2 d \omega}. 
\label{eq:MSEmodel}
\end{equation}
Figure~\ref{fig:tradeoff} shows the trade-off achieved by constrained designs for for \armaxtext{}-1.  (To obtain a meaningful summary of the fluctuating perturbation at the end of the experiment, we used the peak value of $\mathbb{E}[|\yd_{\ts}|]$ during the final second.) We see that the adaptive probing signal matches the performance of \prbstext{}, but effectively controls the resulting perturbations.

\subsection{Effect of delays}

Next, we consider the following system (\armaxtext{}-2):
\begin{equation}
    \begin{split}
        A(q) &= 1 - 1.5 q^{-1} + 0.7 q^{-2}\\  
        B(q) &= 0.5 q^{-1} + 0.1 q^{-2}\\
        C(q) &= 1 + 0.3 q^{-1},
    \end{split}\label{armaxcase2} 
\end{equation}
both without and with an input delay $(\orderdelay=0$ and $\orderdelay=3$, respectively). In this case the output disturbance has a standard deviation $\sigma_v \approx  0.038$. 
The system is regulated by a PI-controller
\begin{equation}
    K(q) = \frac{0.002+0.002194 q^{-1}}{1 - q^{-1}}.
\end{equation}

\begin{figure*}[!th]
   \centering
\subfloat{\includegraphics[width=0.41\linewidth]{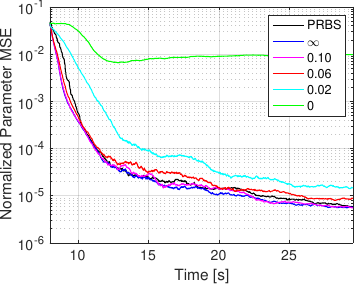}\label{fig:armax-pi-mse-para_2}}\subfloat{\includegraphics[width=0.41\linewidth]{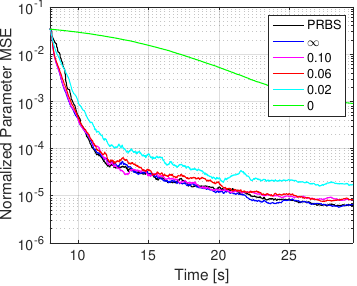}\label{fig:armax-pi-mse-para_2_delay_nk3}}
    \caption{Mean squared error $\text{MSE}_t$ \eqref{eq:MSEparameter} for \armaxtext{}-2 when using \prbstext{} and the designed probing signals $\vd_{\ts}$ for various constraints 
$\ydmax$. (a): With no extra delay $\orderdelay=0$. (b): With extra delay $\orderdelay = 3$.}   \vspace{-0.5em}
\end{figure*}
For this system, a delay of \(\orderdelay=0\) or \(\orderdelay=3\) yields \(\gamma=2\) or \(\gamma=-1\), respectively. This means that a probing signal with persistent excitation of any order is sufficient in the first case while zero probing is required in the latter case. 

We consider $\dmax=0.05$ and vary $\ydmax$. Since  \(\nu \approx 5.14\) in  \eqref{eq:extremal_nu}, any limit $\ydmax>0.26$ is sufficient for recursive feasibility. This result is corroborated by Figure~\ref{fig:zero_ratio_sys2} in Appendix~\ref{ap:result-armax-2} which shows the (estimated) probability of $d_t=0$, which occurs when the design problem is infeasible.

Comparing Figures~\ref{fig:armax-pi-mse-para_2} and ~\ref{fig:armax-pi-mse-para_2_delay_nk3} confirm that the additional delay of $\orderdelay=3$ enables  identifying the system model with zero probing. However, the parameter convergence is dramatically faster when allowing even a small perturbation $\ydmax=0.02$. Since the nonzero delay $\orderdelay$ is inferred from the expanded parameter estimate $\paraexpest_{\ts-1}$ and applied in the constraint \eqref{eq:designproblem}, we expect a larger initial perturbation when $\orderdelay>0$. This effect is observed in the simulations; additional perturbation comparisons are given in Appendix~\ref{ap:result-armax-2}, Figure.~\ref{fig:perturbationmagnitude_2}--\ref{fig:perturbationmagnitudefirst010_2_nk3}. 
\subsection{Frequency Domain Characteristics}
\begin{figure}[!th]
    \centering
    \includegraphics[width=0.82\linewidth]{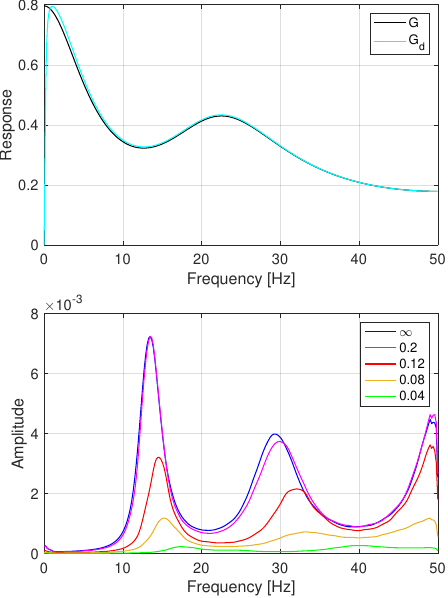}
    \caption{\armaxtext{}-1. (a) Frequency response  $|G(\omega)|^2$ and sensitivity function $|G_\vd(\omega)|^2$. (b) Power spectrum of designed $\vd_{\ts}$ for varying constraints $\ydmax$.}
    \label{fig:spectrum-case1} 
\end{figure}

For \armaxtext{}-1, the frequency response and sensitivity functions are illustrated in Figure~\ref{fig:spectrum-case1}, where one can observe two peaks located around 0 and 22 Hz. We study the spectra of the probing signal $\vd_{\ts}$ as $\ydmax$ is varied. We observe in Figure ~\ref{fig:spectrum-case1} that when $\ydmax$ decreases, not only does the power of $\vd_{\ts}$ decrease but it is spread to the frequencies where the sensitivity function has smaller gains.
A corresponding spectral comparison for \armaxtext{}-2 is provided in Appendix~\ref{ap:result-armax-2} Figure~\ref{fig:spectrum-case2}.

\section{Conclusion}
We have studied closed-loop \armaxtext{} identification under output feedback.
The analysis shows that  identifiability are guaranteed
when the controller order and input delay are sufficiently large relative
to the \armaxtext{} model orders. Conversely, when this order condition is not satisfied,
the missing informativeness can be supplied by a bounded, discrete-valued,
sufficiently rich probing signal, without requiring independence between
the probing signal and the innovation. 

We then derived a closed-form recursive experiment design that generates
such probing signals online. The design keeps the probing 
informative while steering the resulting output perturbation toward
prescribed limits, unlike standard \prbstext{} designs. Moreover, we
showed that there exist perturbation limits under which the constrained
probing design is recursively feasible.

\bibliographystyle{abbrv} 
\bibliography{main}
\appendix

\section{Proof of Proposition~\ref{prop:recursive_feasibility}}\label{ap:proof-recursive-feasiblity}
\begin{pf}
Consider an arbitrary perturbation that is bounded in $[\underline{\delta},\overline{\delta}]$. This can be expressed as
\begin{equation}
\underline{\delta} \leq \delta_{\ts+\orderdelay} = g_\K^\top D_t \leq \overline{\delta},
\label{eq:proof_basecase}
\end{equation}
where 
\[
D_t \triangleq [\vd_{\ts-1}\; \vd_{\ts-2}\; \cdots\; \vd_{\ts-\K+\orderdelay}]^\top .
\]
We want to derive limits $\underline{\delta} \leq 0 \leq \overline{\delta}$ in which 
\begin{equation}
\underline{\delta} \leq \delta_{\ts+\orderdelay+1}
=\widetilde{g}_{\orderdelay+1} \vd_{\ts} + \constb_{\ts} \leq \overline{\delta}
\label{eq:proof_recursivefeasible}
\end{equation}
is feasible.

By rearranging \eqref{eq:proof_recursivefeasible}, we see that it is sufficient to show the following inequalities,
\begin{equation}
\underline{\delta}  - |\constA|\dmax \le \constb_\ts
\le
\overline{\delta}+|\constA|d_{\max},
\label{eq:proof_recursivefeasible_sufficient}
\end{equation}
hold simultaneously. Note that $\constb_\ts$ in \eqref{eq:h_history} collects the contribution of past probing signals and can be expressed as
\[
\constb_\ts=s_\K^\top D_\ts.
\]


We now turn to deriving upper and lower bounds on $\constb_\ts$, by writing 
\[s_\K^\top D = (s_k-v g_\K)^\top D + v g_\K^\top D,\; v\in[0,1).\]
We begin with the upper bound. First, we have that 
\begin{equation}
\begin{split}
(s_k-v g_\K)^\top D &\le \|s_k-v g_k\|_1\|D\|_\infty \\
&\leq \|s_k-v g_k\|_1\dmax.
\end{split}\label{eq:recursivefeasiblity-norm}
\end{equation}
using the H\"older inequality. Then by \eqref{eq:proof_basecase}, we have
\begin{equation}
 v g_\K^\top D\le v \overline{\delta}\label{eq:recursivefeasiblity-holder}\end{equation}
By combining \eqref{eq:recursivefeasiblity-norm} and \eqref{eq:recursivefeasiblity-holder}, the inequality
\[s_\K^\top D \le \|s_k-v g_k\|_1\dmax+ v \overline{\delta}\]
holds for all perturbations satisfying \eqref{eq:proof_basecase}. From the second inequality in \eqref{eq:proof_recursivefeasible_sufficient}, we seek a nonnegative $\overline{\delta}$ that satisfies
\begin{equation}
\|s_k-v g_k\|_1\dmax+ v \overline{\delta} \leq 
\overline{\delta}+|\constA|d_{\max},
\end{equation}
that is,
\[\overline{\delta}\ge \dmax \frac{\|s_k-v g_k\|_1-|\constA|}{1-v}.\]
The tightest upper bound is thus 
\[\overline{\delta}\ge \dmax \nu,\]
given by \eqref{eq:extremal_nu}.

Similarly, to address the first inequality in \eqref{eq:proof_recursivefeasible_sufficient}, we use the fact that
\[(s_k-v g_\K)^\top D\ge -\|s_k-v g_k\|_1\|D\|_\infty,\;v\in [0,1). \]
to obtain
\[\underline{\delta}\le -\dmax\frac{\|s_k-v g_k\|_1-|\constA|}{1-v}.\]
The tightest lower bound is thus 
\[\underline{\delta}\le -\dmax \nu.\]
\end{pf}
\section{Algorithmic summary}\label{ap:alg1}
\begin{algorithm}[!th]
\caption{Recursive Design Method}
\label{alg:sequential-probing}
\begin{algorithmic}[1]
\State Start the algorithm at $t_0$
\For{$\ts =\ts_0,\ts_0+1,\ldots$}
    \State Apply $\vpinput_t=\vinput_\ts+\vd_\ts$ and collect $(\vpinput_t,\voutput_\ts)$.
    \State Update the recursive estimates $\paraexpest_\ts$, $P_t$ using the newly collected data.
    \State Estimate $\widehat{n}_{d,t}$
    as in \eqref{eq:estimate_delay}.
    \State Compute the
    $\widetilde{g}_{\widehat{n}_{d,t}+1}(\paraest_t)$ and $h_t(\paraest_t)$
    from \eqref{eq:finitewindowperturb}.
    \State Compute design quantities $R_t,\xi_t$ from \eqref{eq:rmatrix}.
    \State Compute the feasible interval $[d_{l,t},d_{u,t}]$ from \eqref{eq:dl} and \eqref{eq:du}.
    \State Determine the next probing input $d_{t+1}$ subject to $[d_{l,t},d_{u,t}]$ by \eqref{eq:solution-siso}.
\EndFor
\end{algorithmic}
\end{algorithm}

\section{Additional result for \armaxtext{}-2}\label{ap:result-armax-2}
\begin{figure}[!thb]
    \centering    \includegraphics[width=0.82\linewidth]{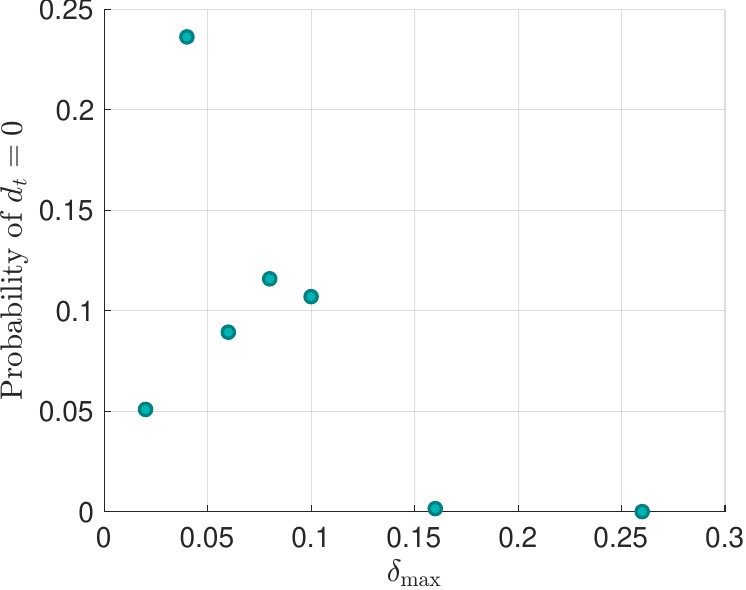}
    \caption{Probability of infeasible design (so that $d_t=0$) with respect to $\ydmax \in\{0.02,0.04,0.06,0.08,0.1,0.16,0.26\}$ for \armaxtext{}-2 \eqref{armaxcase2} with $\orderdelay=0$.}
    \label{fig:zero_ratio_sys2}
\end{figure}

\begin{figure*}[!th]
    \centering
    \subfloat{\includegraphics[width=0.41\linewidth]{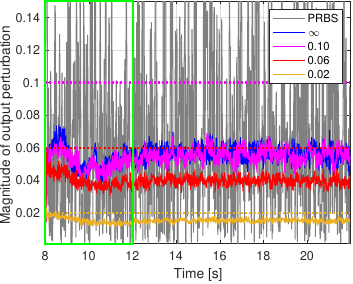}\label{fig:perturbationmagnitude_2}}\subfloat{\includegraphics[width=0.41\linewidth]{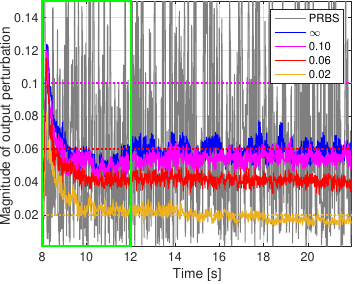}\label{fig:perturbationmagnitude_2_delay_nk3}}
    \caption{Magnitude of output perturbation $\mathbb{E}[|\yd_{\ts}|]$ for\armaxtext{}-2 when using \prbstext{} and the designed probing signals. Dotted lines show user-specified limits $\ydmax$. (a): With no extra delay $\orderdelay=0$. (b): With extra delay $\orderdelay = 3$.}  
\end{figure*}

\begin{figure*}[!th]
\centering
\subfloat{\includegraphics[width=0.41\linewidth]{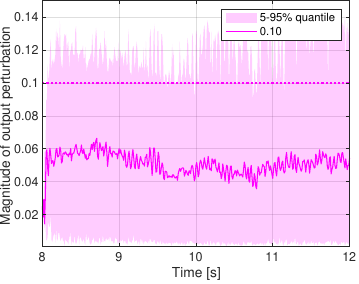}\label{fig:perturbationmagnitudefirst010_2}}\subfloat{\includegraphics[width=0.41\linewidth]{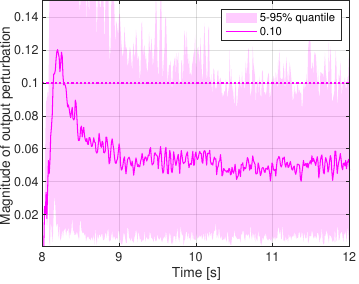}\label{fig:perturbationmagnitudefirst010_2_nk3}}
    \caption{Zoomed-in view (first 400 samples) of the output perturbation magnitude corresponding to the boxes in Fig.~\ref{fig:perturbationmagnitude_2} and ~\ref{fig:perturbationmagnitude_2_delay_nk3}, $\ydmax=0.1$. The solid line represents the mean over 100 Monte Carlo runs, and the shaded region shows the 5--95\% quantile band. (a): \(\orderdelay=0\). (b): \(\orderdelay=3\).}  
\end{figure*}
\begin{figure}[!thb]
    \centering
    \includegraphics[width=0.82\linewidth]{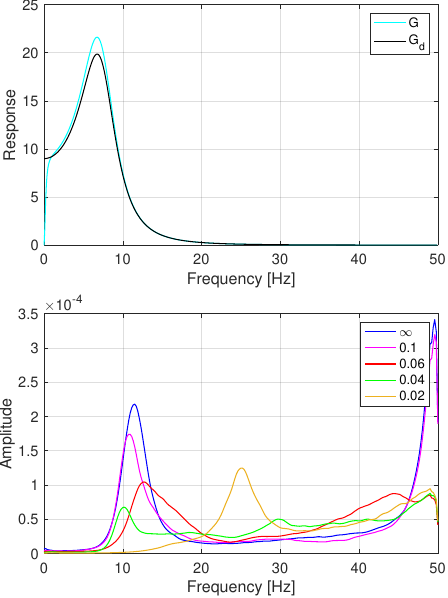}
    \caption{\armaxtext{}-2. (a) Frequency response $|G(\omega)|^2$ and sensitivity function $|G_\vd(\omega)|^2$. (b) Power spectrum of designed perturbation.}
    \label{fig:spectrum-case2}
\end{figure}

\end{document}